\documentstyle[aps]{revtex}
\begin{document}
\centerline{\LARGE\bf We can know more in double-slit experiment}
\vskip 1cm \centerline{Gao Shan} \centerline{Institute of Quantum
Mechanics} \centerline{11-10, NO.10 Building, YueTan XiJie DongLi,
XiCheng District} \centerline{Beijing 100045, P.R.China}
\centerline{E-mail: gaoshan.iqm@263.net}

\vskip 1cm
\begin{abstract}
We show that we can know more than the orthodox view does, as one
example, we make a new analysis about double-slit experiment, and
demonstrate that we can measure the objective state of the
particles passing through the two slits while not destroying the
interference pattern, the measurement method is to use protective
measurement.

\end{abstract}

\vskip 1cm

Double-slit experiment has been widely discussed, and nearly all
textbooks about quantum mechanics demonstrated the weirdness of
quantum world using it as one example, as Feynman said, it
contains all mysteries of quantum mechanics, but have we disclosed
these mysteries and understood the weirdness in double-slit
experiment? as we think, the answer is definitely No.

When discussing double-slit experiment, the most notorious
question is which slit the particle passes through in each
experiment, it is just this problem that touches our sore spots in
understanding quantum mechanics, according to the widely-accepted
orthodox view, this question is actually meaningless, let's see
how it gets this bizarre answer, it assumes that only an
measurement can give an answer to the above question, then
detectors need to be put near both slits to measure which slit the
particle passes through, but when this is done the interference
pattern will disappear, thus the orthodox view asserts that the
above question is meaningless since we can not measure which slit
the particle passes through while not destroying the interference
pattern.

In fact, the above question is indeed meaningless, and at it
happens the orthodox answer is right, but its reason is by no
means right, the genuine reason is that if the particle passes
through only one slit in each experiment, the interference pattern
will not be formed at all\footnote{Here we assume the only
existence of particle, thus Bohm's hidden-variable
theory\cite{Bohm} is not considered.}, thus it is obviously wrong
to ask which slit the particle passes through in each experiment,
it does not pass through a single slit at all!

On the other hand, we can still ask the following meaningful
question, namely how the particle passes through the two slits to
form the interference pattern? now as to this question, the deadly
flaw of the orthodox view is clearly unveiled, what is its answer?
as we know, its answer will be there does not exist any objective
motion picture of the particle, the question is still meaningless,
but how can it get this conclusion? it can't! and no one can.

Since we have known that the particle does not pass through a
single slit in each experiment, the direct position measurement
near both slits is obviously useless for finding the objective
motion state of the particle passing through the two slits, and it
will also destroy the objective motion state of the particle, then
the operational basis of the orthodox view disappears, it also
ruins, thus the orthodox demonstrations can't compel us to reject
the objective motion picture of the particle\footnote{Why we can't
detect which slit the particle passes through when not destroying
the interference pattern is not because there does not exist any
objective motion picture of the particle, but because the particle
does not pass through a single slit at all.}, it only requires
that the motion picture of classical continuous motion should be
rejected, this is undoubtedly right, since the motion of
microscopic particle will be not classical continuous motion at
all, it will be one kind of completely different motion.

Once the objective motion picture of the particle can't be
essentially rejected, we can first have a look at it using the
logical microscope, since the particle does not pass through a
single slit in each experiment, it must pass through both slits
during passing through the two slits, it has no other choices!
this kind of bizarre motion is not impossible since it will take a
period of time for the particle to pass through the slits, no
matter how short this time interval is, so far as it is not zero,
the particle can pass through both slits during this finite time
interval, what it must do is just discontinuously move, nobody can
prevent it from moving in such a way! in fact, as we have
demonstrated\cite{Gao}, this is just the natural motion of
particle.

On the other hand, in order to find and confirm the objective
motion picture of the particle passing through the two slits,
which will be very different from classical continuous motion, we
still need a new kind of measurement, which will be very different
from the position measurement, fortunately it has been found
several year ago\cite{Aha1,Aha2}, its name is protective
measurement, since we know the state of the particle beforehand in
double-slit experiment, we can protectively measure the objective
motion state of the particle when it passes through the two slits,
while the state of the particle will not be destroyed after such
protective measurement, and the interference pattern will not be
destroyed either, thus by use of this kind of measurement we can
find the objective motion picture of the particle passing through
the two slits while not destroying the interference pattern, and
the measurement results will reveal that the particle indeed
passes through both slits as we see using the logical microscope.

Now, the above analysis has strictly demonstrated that we can know
more than the orthodox view does in double-slit experiment, namely
we know that the particle passes through both slits to form the
interference pattern, while the orthodox view never knows this.

\end{document}